\documentclass[aps,prl,twocolumn,floats,epsfig,showpacs]{revtex4-1}

\usepackage{bbm}
\usepackage{amssymb}
\usepackage{ifpdf}
\usepackage{color}
\usepackage{graphicx,epsfig}

\newcommand{\Z}{{\mathbb{Z}}}

\newcommand{\RP}{{\mathbb{R}P}}

\begin{document}

\title{The $(2+1)$-d $U(1)$ Quantum Link Model Masquerading as Deconfined 
Criticality$^*$}
\author{D.\ Banerjee$^1$, F.-J.\ Jiang$^2$, P.\ Widmer$^1$, and 
U.-J.\ Wiese$^{1,3}$}
\affiliation{$^1$Albert Einstein Center, Institute for Theoretical Physics, 
Bern University, Switzerland \\
$^2$ Department of Physics, National Taiwan Normal University
88, Sec.\ 4, Ting-Chou Rd., Taipei 116, Taiwan \\
$^3$ Center for Theoretical Physics, Massachusetts Institute of Technology,
Cambridge, Massachusetts, U.S.A. \\
$^*$ {\bf Dedicated to the memory of Bernard B.\ Beard (1957-2012)}}

\begin{abstract}
The $(2+1)$-d $U(1)$ quantum link model is a gauge theory, amenable to quantum
simulation, with a spontaneously broken $SO(2)$ symmetry emerging at a quantum 
phase transition. Its low-energy physics is described by a $(2+1)$-d $\RP(1)$ 
effective field theory, perturbed by a dangerously irrelevant $SO(2)$ breaking 
operator, which prevents the interpretation of the emergent pseudo-Goldstone 
boson as a dual photon. At the quantum phase transition, the model mimics some 
features of deconfined quantum criticality, but remains linearly confining. 
Deconfinement only sets in at high temperature.
\end{abstract}

\maketitle

Quantum link models (QLMs) are lattice gauge theories formulated in terms of 
discrete quantum degrees of freedom. $U(1)$ and $SU(2)$ QLMs were first 
constructed by Horn in 1981 \cite{Hor81}, and further investigated in 
\cite{Orl90}. In \cite{Cha97} QLMs were introduced as an 
alternative non-perturbative regularization of Abelian and non-Abelian gauge 
theories, in which ordinary gauge fields emerge dynamically from the 
dimensional reduction of discrete quantum link variables. Dimensional reduction 
of discrete variables is a generic phenomenon in asymptotically free theories,
which gives rise to the D-theory formulation of quantum field theory 
\cite{Bro04}. In the D-theory formulation of 4-d Quantum Chromodynamics (QCD), 
the confining gluon field emerges by dimensional reduction from a deconfined 
Coulomb phase of a $(4+1)$-d $SU(3)$ QLM \cite{Bro99}. Chiral quarks arise 
naturally as domain wall fermions located at the two 4-d sides of a $(4+1)$-d 
slab. The $(2+1)$-d $U(1)$ QLM has also been investigated in the context of 
quantum spin liquids \cite{Her04}. With staggered background charges $\pm 1$,
it is equivalent to a quantum dimer model \cite{Rok88,Moe02,Ral08}. Furthermore,
Kitaev's toric code \cite{Kit06} is a $\Z(2)$ QLM. In contrast to Wilson's 
lattice gauge theory \cite{Wil74}, QLMs have a finite-dimensional Hilbert space 
per link, which makes them ideally suited for the construction of atomic quantum
simulators for dynamical Abelian \cite{Bue05,Zoh12,Ban12,Tag12,Zoh13a} and 
non-Abelian gauge theories \cite{Ban13,Zoh13b,Tag12a,Zoh13c}. A long-term goal 
of this research is to quantum simulate QCD in the D-theory formulation with 
ultracold matter, in order to address the real-time evolution of strongly 
interacting systems in nuclear and particle physics, as well as their dynamics 
at non-zero baryon density.

In this paper, we investigate the $(2+1)$-d $U(1)$ QLM, in order 
to demonstrate that, despite its structural simplicity, it displays highly 
non-trivial dynamics, and thus is ideally suited to demonstrate the power of 
gauge theory quantum simulators. We consider the model with a plaquette 
coupling $J$ and a Rokhsar-Kivelson (RK) coupling $\lambda$. The phase diagram
is sketched in Fig.1. 
\begin{figure}[tbp]
\includegraphics[width=0.35\textwidth]{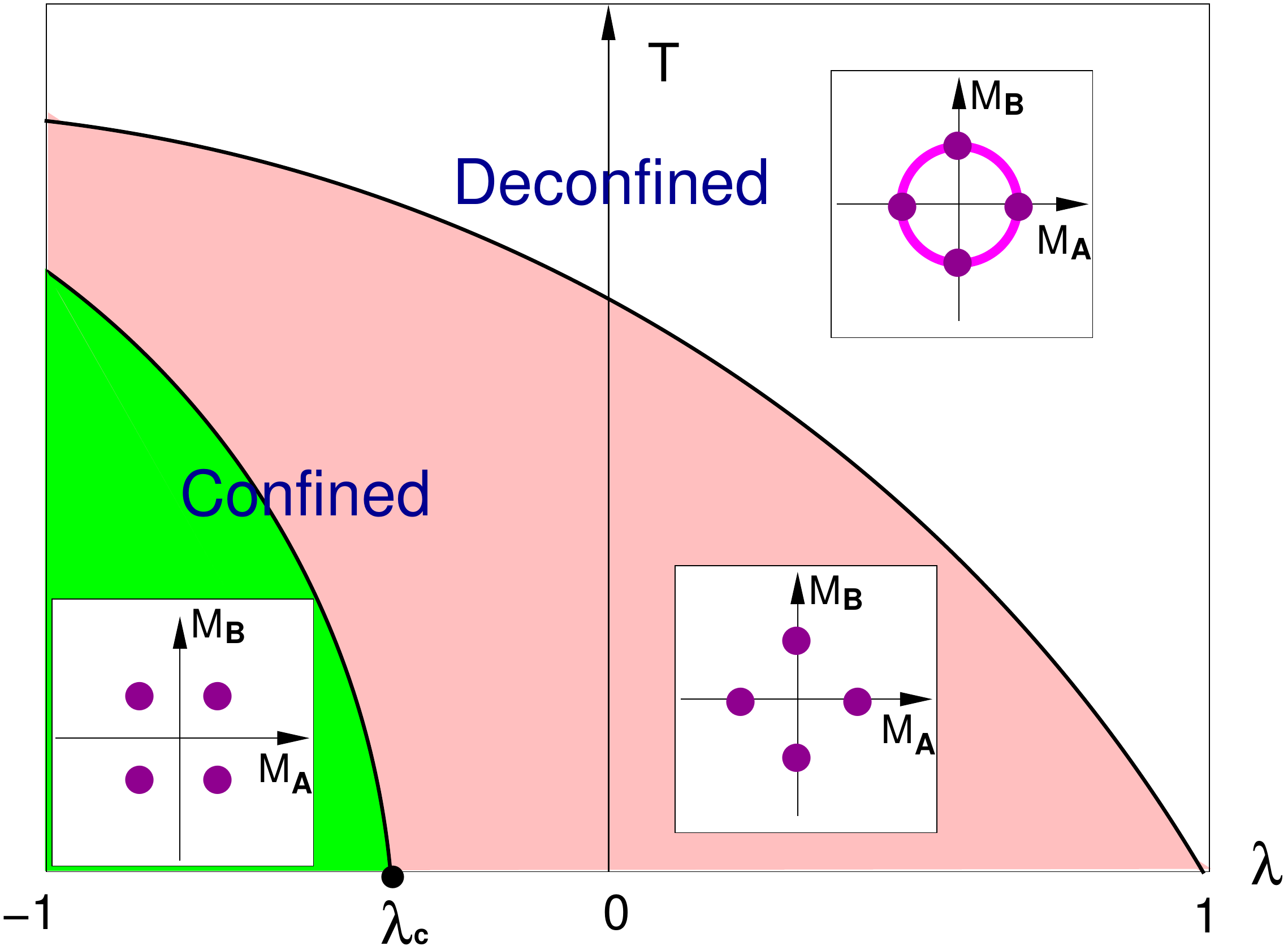}
\caption{[Color online] \textit{Schematic sketch of the $\lambda$-$T$ phase 
diagram. The insets indicate the location of the peaks in the probability
distribution of the order parameter $p(M_A,M_B)$.}}
\end{figure}
At zero temperature, the model is confining for $\lambda < 1$. At finite 
temperature $T$, it has a deconfinement phase transition above which there is a 
massless mode transforming non-trivially under the $U(1)$ center symmetry. Due
to the Mermin-Wagner theorem, this is not associated with spontaneous symmetry 
breaking. The deconfinement phase transition reaches zero temperature at the RK 
point, $\lambda = 1$. At $\lambda_c$ there is a quantum phase transition 
which separates two phases with spontaneously broken translation symmetry 
\cite{Sha04}. The phase at $\lambda < \lambda_c$ has, in addition, a 
spontaneously broken charge conjugation symmetry. The two phases are similar to 
the columnar and plaquette ordered valence bond solid phases in a quantum dimer 
model \cite{Moe02}, which may be separated by a first order phase transition 
\cite{Vis04} or by an intermediate phase \cite{Ral08}.

As we will see, at $\lambda_c$ a spontaneously broken approximate global 
$SO(2)$ symmetry emerges dynamically, giving rise to a light pseudo-Goldstone 
boson. The interface that separates the two broken phases on either side of the 
transition manifests itself as a string with fractional electric flux 
$\frac{1}{2}$. This raises the question whether the phase transition might be a 
deconfined quantum critical point in the sense of \cite{Sen04,Sen04a}, 
corresponding to a conformal field theory with an emergent massless photon and 
deconfined 
electric charges. Deconfined quantum criticality has first been investigated 
numerically in the $J$-$Q$ quantum spin model \cite{San07,Mel08,Jia08}, between 
an antiferromagnetic and a valence bond solid phase, and is still discussed 
controversially \cite{Che13,Tan13}. It has also been studied in the $J$-$Q$ 
\cite{Dam13} and in the $J_1$-$J_2$ model \cite{Alb11,Zhu12,Gan13} on the 
honeycomb lattice. At a deconfined quantum critical point, the instanton-like 
monopole events that cause permanent confinement in a $(2+1)$-d compact $U(1)$ 
gauge theory \cite{Pol75,Goe81} are eliminated, because a $\Z(4)$-invariant 
term that explicitly breaks the emergent $SO(2)$ symmetry of the effective 
action becomes irrelevant. We will see that this is not what happens in the 
$(2+1)$-d $U(1)$ QLM, where the $\Z(4)$-invariant term can be tuned to zero. 
Still, the emergent $SO(2)$ symmetry remains weakly explicitly broken by a 
``dangerously irrelevant'' operator \cite{Bru75,Ami82}. This prevents the 
interpretation of the Goldstone boson as an emergent dual photon. It is more 
appropriate to think of it as an accidentally light Abelian ``glueball''. The 
dangerously irrelevant operator also contributes to the string tension and 
implies that the theory remains confining at the phase transition. 

The Hamiltonian of the $(2+1)$-d $U(1)$ QLM is
\begin{equation}
H = - J \sum_{\Box} \left[U_\Box + U_\Box^\dagger - 
\lambda (U_\Box + U_\Box^\dagger)^2\right].
\end{equation}
Here $U_\Box = U_{wx} U_{xy} U_{zy}^\dagger U_{wz}^\dagger$ is a plaquette operator 
formed by quantum links $U_{xy}$ connecting nearest-neighbor sites $x$ and $y$ 
on a 2-d square lattice. A $U(1)$ quantum link $U_{xy} = S_{xy}^+$ is a raising 
operator of electric flux $E = S_{xy}^3$, constructed from a quantum spin 
$\vec S_{xy}$ associated with the link $xy$. In Wilson's lattice gauge theory, 
where the link variables are classical parallel transporters, 
$U_{xy} = \exp(i \varphi_{xy}) \in U(1)$, taking values in the gauge group, and 
$E_{xy} = - i \partial_{\varphi_{xy}}$, the single-link Hilbert space is
infinite-dimensional. In the $U(1)$ QLM, on the other hand, it is just given by 
a finite-dimensional representation of the embedding algebra $SU(2)$. When one 
chooses spin $\frac{1}{2}$ on each link, the link Hilbert space is just 
2-dimensional. The first term in the Hamiltonian 
flips a loop of electric flux, winding around an elementary plaquette, and 
annihilates non-flippable plaquette states, while the RK term, proportional to 
$\lambda$, counts flippable plaquettes. The Hamiltonian commutes with the 
generators, $G_x = \sum_i (E_{x,x+\hat i} - E_{x-\hat i,x})$, of infinitesimal 
$U(1)$ gauge transformations. Here $\hat i$ is a unit-vector pointing in the 
$i$-direction. Physical states $|\Psi\rangle$ are gauge invariant, i.e.\ they
obey the Gauss law $G_x |\Psi\rangle = 0$. Besides the gauge symmetry, the QLM
also has several global symmetries, including lattice 
translation invariance and charge conjugation. Translation invariance 
characterizes each energy eigenstate by its lattice momentum $p = (p_1,p_2) \in 
\ ]- \pi,\pi]^2$. Charge conjugation replaces $U_{xy}$ by $U_{xy}^\dagger$ and 
$E_{xy}$ by $- E_{xy}$, and characterizes each eigenstate by its charge 
conjugation parity $C = \pm$. Another important global symmetry is the center 
symmetry associated with ``large'' gauge transformations. The $U(1)$ QLM 
defined on a periodic volume has super-selection sectors characterized by 
wrapping electric fluxes that take values in $\Z/2$. On an $L_1 \times L_2$ 
lattice with periodic boundary conditions, the generators of the $U(1)$ 
center symmetry are $E_i = \frac{1}{L_i} \sum_x E_{x,x+\hat i}$. They commute with 
the Hamiltonian, but cannot be expressed through ``small'' periodic gauge 
transformations $G_x$.

We have performed exact diagonalization studies of the $(2+1)$-d $U(1)$ QLM 
with $S = \frac{1}{2}$ on $4 \times 4$, $4 \times 6$, and 
$6 \times 6$ lattices. The energies of the lowest states are illustrated in 
Fig.2a. For $\lambda < 1$, the ground state has 
momentum $(0,0)$ and is even under charge conjugation (i.e.\ $C = +$). For 
$\lambda < \lambda_c$ the first excited state has quantum numbers $C = -$, 
$p = (\pi,\pi)$. Its energy gap to the ground state, 
$E_- \sim \exp(- \sigma_- L_1 L_2)$, decreases exponentially with the 
volume $L_1 L_2$, thus indicating the spontaneous breakdown of charge 
conjugation C and the translation T by one lattice spacing (in either 
direction). For $\lambda > \lambda_c$, another state 
$|C=+,p = (\pi,\pi)\rangle$ degenerates with the ground state, i.e.\ 
$E_+ \sim \exp(- \sigma_+ L_1 L_2)$, indicating that C is now restored, 
while T remains spontaneously broken. The crossing of the two excited energy 
levels, $E_-(\lambda_{pc}) = E_+(\lambda_{pc})$, defines a volume-dependent 
pseudo-critical coupling $\lambda_{pc}$. The next zero-momentum excited states, 
$|C=\pm,p = (0,0)\rangle$ with energies $E'_{\pm}$, cross twice near the critical
point at two pseudo-critical couplings ${\lambda_{pc}'}^{\!\!\!+}$ and 
${\lambda_{pc}'}^{\!\!\!-}$, i.e.\ $E'_-({\lambda'_{pc}}^{\!\!\!\pm}) = 
E'_+({\lambda'_{pc}}^{\!\!\!\pm})$. As illustrated in Fig.2b, the couplings 
$\lambda_{pc}$, ${\lambda'_{pc}}^{\!\!\!+}$, and ${\lambda'_{pc}}^{\!\!\!-}$ all 
approach $\lambda_c = - 0.359(5)$ in the infinite volume limit.
\begin{figure}[tbp]
\includegraphics[width=0.5\textwidth]{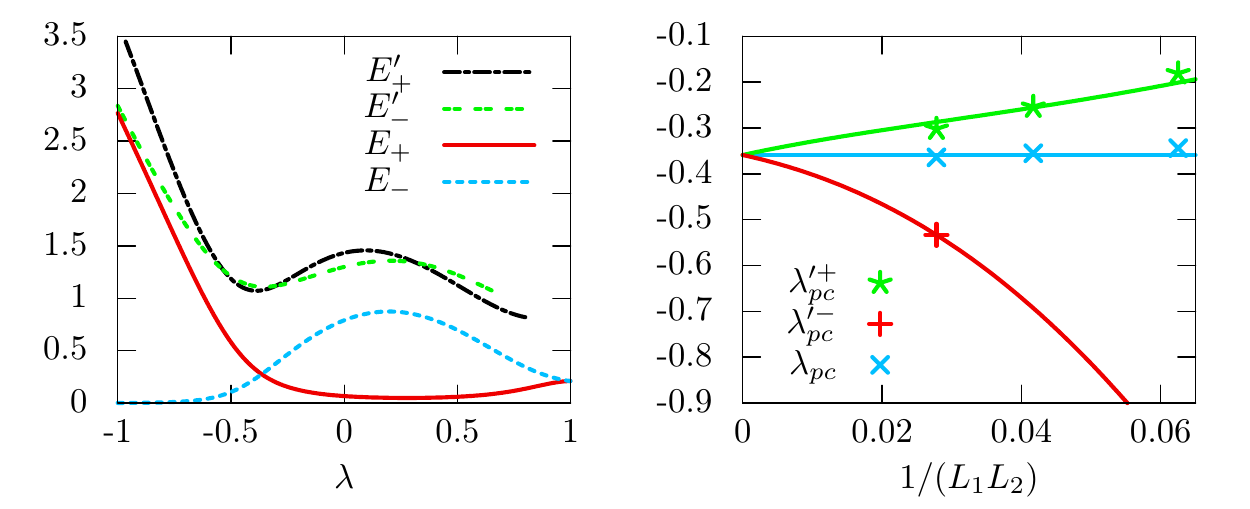}
\caption{[Color online] \textit{a) Energy gaps of the lowest states on the 
$6 \times 6$ lattice as a function of $\lambda$. b) Global fit of 
$L_1 L_2$-dependence of $\lambda_{pc}$, ${\lambda'_{pc}}^{\!\!\!+}$, and 
${\lambda'_{pc}}^{\!\!\!-}$, that yields $\lambda_c = -0.359(5)$.}}
\end{figure}

The different symmetry breaking patterns are distinguished by two order 
parameters, $M_A$ and $M_B$, associated with the even and odd dual sublattices 
$A$ and $B$. A configuration of quantum height variables 
$h^A_{\widetilde x} = 0,1$, $h^B_{\widetilde x} = \pm \frac{1}{2}$, located at the 
dual sites $\widetilde x = (x_1 + \frac{1}{2},x_2 + \frac{1}{2})$, is associated
with a flux configuration $E_{x,x+\hat i} =
[h^X_{\widetilde x} - h^{X'}_{\widetilde x+ \hat i - \hat 1 - \hat 2}] \mbox{mod} 2
= \pm \frac{1}{2}$, $X,X' \in \{A,B\}$. The two order parameters are given by
$M_X = \sum_{\widetilde x \in X} s^X_{\widetilde x} h^X_{\widetilde x}$, where
$s^A_{\widetilde x} = (-1)^{(\widetilde x_1 - \widetilde x_2)/2}$ and 
$s^B_{\widetilde x} = (-1)^{(\widetilde x_1 - \widetilde x_2 + 1)/2}$. Under C and T 
they transform as $^C M_A = M_A$, $^C M_B = - M_B$, $^T M_A = - M_B$, 
$^T M_B = M_A$. It should be pointed out that $\pm (M_A,M_B)$ represent the same 
physical configuration, because shifting the height variables to 
$h^X_{\widetilde x}(t)' = [h^X_{\widetilde x}(t) + 1] \mbox{mod} 2$
leaves the electric flux configuration unchanged. We have performed quantum 
Monte Carlo simulations with an efficient newly developed cluster algorithm,
that will be described elsewhere. The algorithm has been used to determine the 
probability distribution $p(M_A,M_B)$ of the two order parameters $M_A$ and 
$M_B$ shown in Fig.3 at $\lambda = -1$, $\lambda_c$, and 0 for 
$L_1 = L_2 = 24 a$, which reveals an emergent spontaneously broken $SO(2)$ 
symmetry at the quantum phase transition.
\begin{figure}[tbp]
\includegraphics[width=0.23\textwidth]{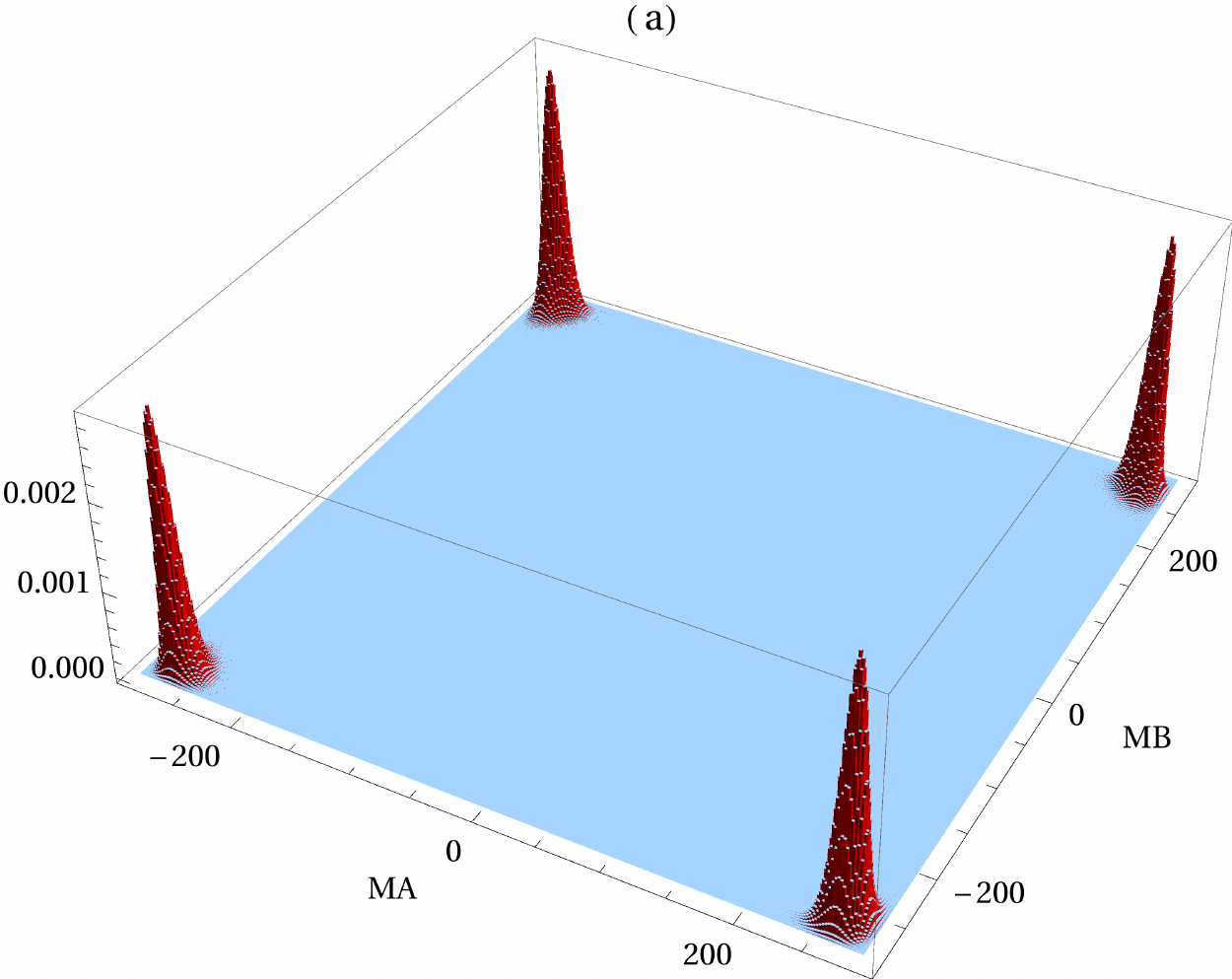}
\includegraphics[width=0.23\textwidth]{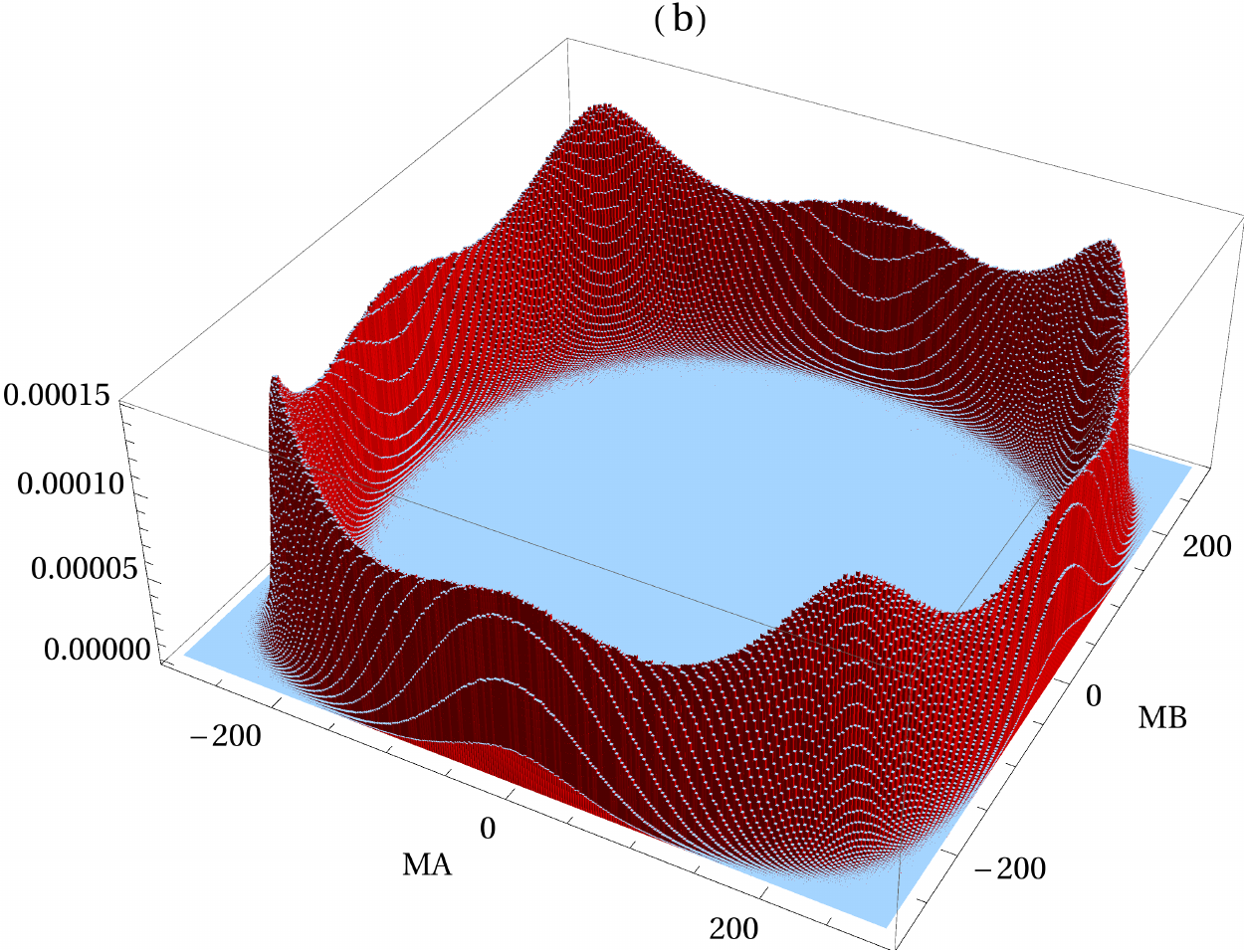}
\includegraphics[width=0.23\textwidth]{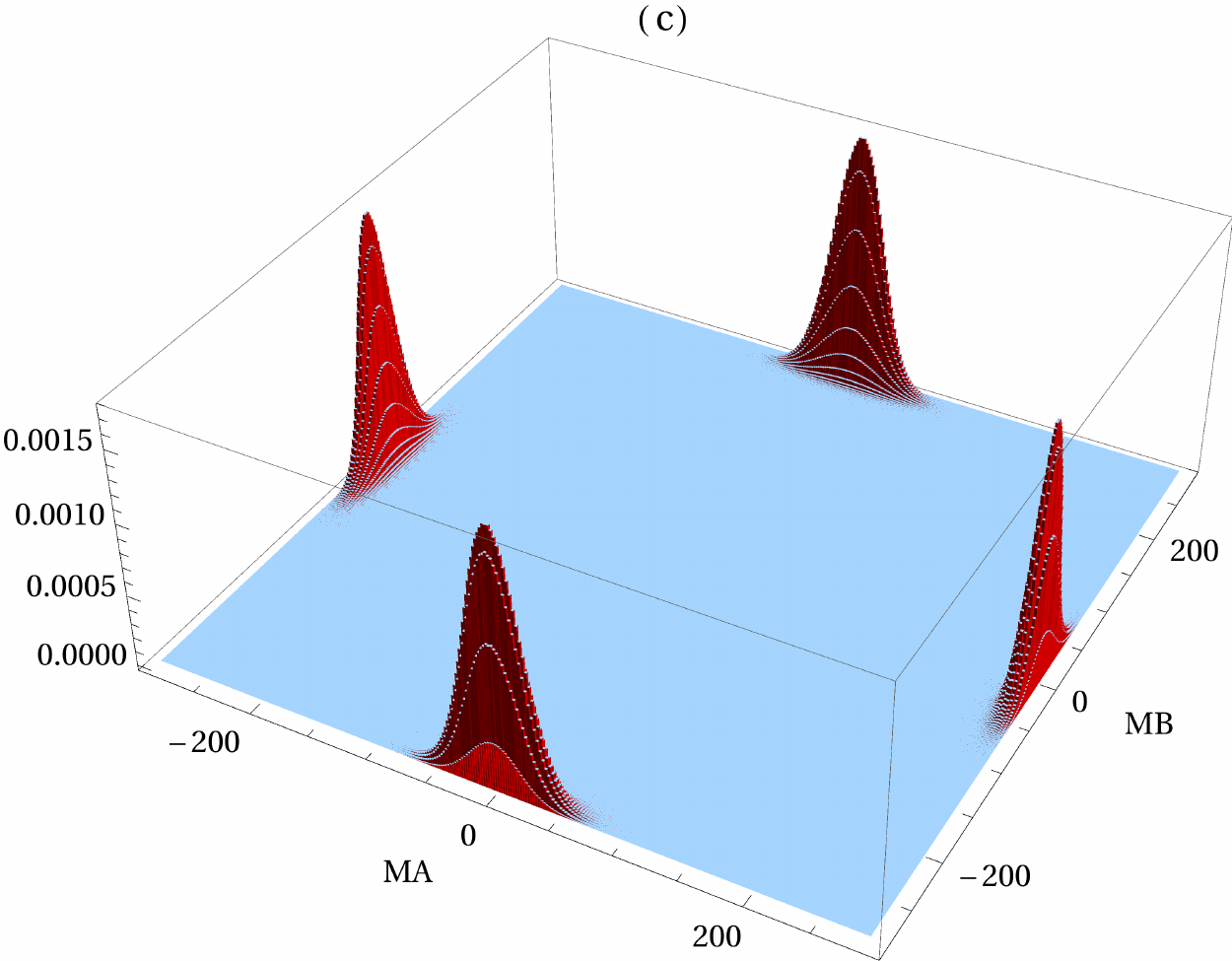}
\includegraphics[width=0.23\textwidth]{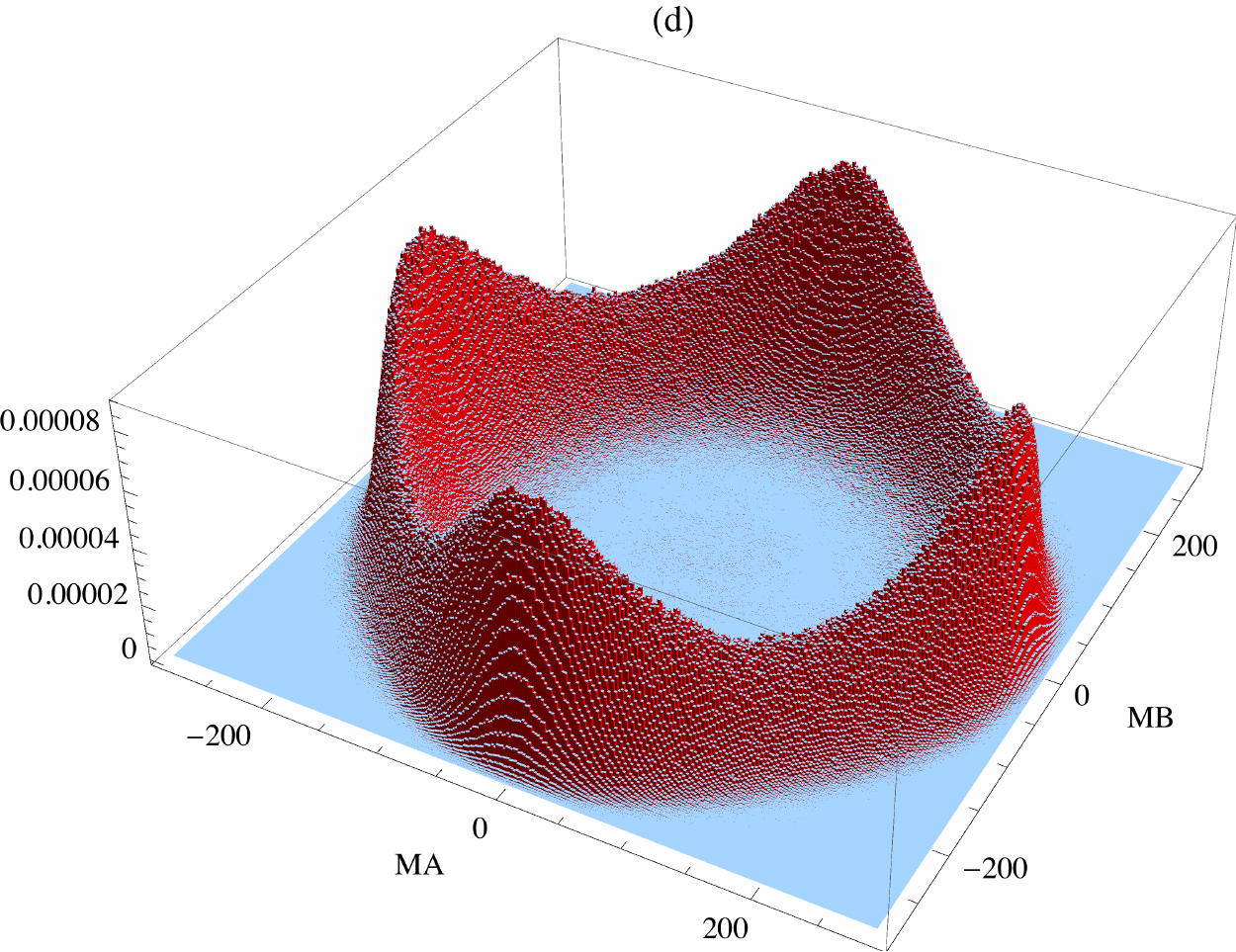}
\caption{[Color online] \textit{Probability distribution $p(M_A,M_B)$ for
$\lambda = -1$, $\lambda_c$, and 0 at $T = 0$ (a,b,c), and $\lambda = 0$, 
$T>T_c$ (d).}}
\end{figure}

The low-energy effective theory describing the vicinity of the quantum phase 
transition is formulated in terms of a unit-vector field 
$\vec e(x) = (\cos\varphi(x),\sin\varphi(x))$ representing the direction of
$(M_A,M_B)$. Since $(M_A,M_B)$ and $- (M_A,M_B)$ are indistinguishable, the
effective theory is a $(2+1)$-d $\RP(1)$ model. Thus, only those states that 
are invariant against a sign-change of $\vec e(x)$ belong to the physical 
Hilbert space. Introducing $\partial_3 = \partial_{ct}$, the corresponding 
Euclidean effective action is
\begin{equation}
S[\varphi] \! = \! \int \! d^3x \frac{1}{c} \! \left[
\frac{\rho}{2} \partial_\mu \varphi \partial_\mu \varphi \! + \! 
\delta \cos^2(2 \varphi) \! + \! \varepsilon \cos^4(2 \varphi)\right].
\end{equation}
Here $\rho$ is the spin stiffness and $c$ is the velocity of an emergent 
pseudo-Goldstone boson. $\delta + \varepsilon$ measures the deviation from the 
phase transition. The $\delta$-term explicitly breaks the emergent $SO(2)$ 
symmetry to a $\Z(4)$ subgroup and gives rise to a small Goldstone boson mass 
$M c = 2 \sqrt{2|\delta|/\rho}$. Even when the relevant $\delta$-term is tuned 
to zero, the dangerously irrelevant $\varepsilon$-term still explicitly breaks 
the $SO(2)$ symmetry. It is natural to define the dual field
$F_{\mu\nu}(x) = \frac{1}{\pi} \varepsilon_{\mu\nu\rho} \partial_\rho \varphi(x)$. 
Since $\varphi(x)$ is well-defined only up to multiples of $\pi$, vortices and 
half-vortices in the order parameter field manifest themselves as charges. The 
electric charge contained in a spatial region $\Omega$ is given by twice the 
vortex number
\begin{equation}
Q_\Omega = \int_\Omega d^2x \ \partial_i F_{0i} = 
\frac{1}{\pi} \int_{\partial \Omega} d\sigma_i \ 
\varepsilon_{ij} \partial_j \varphi \in \frac{\Z}{2}.
\end{equation}
Note that a charge 1 corresponds to a half-vortex, which is allowed because 
$\vec e(x)$ and $- \vec e(x)$ are physically equivalent. While the flux of 
$F_{\mu\nu}$ correctly represents the conserved charges of the $U(1)$ center 
symmetry, $F_{\mu\nu}$ should not be mistaken for a dual massless photon. This 
interpretation would require an exact $SO(2)$ symmetry, at least in the 
infrared. Due to the $\varepsilon$-term and other higher order symmetry 
breaking terms, this would require a large amount of fine-tuning.

By applying the Ginsburg-Landau-Wilson paradigm to the $\delta$- and
$\varepsilon$-terms, in mean field theory one obtains the phase diagram of 
Fig.4. The two phases realized in the QLM both have four peaks 
in the order parameter distribution $p(M_A,M_B)$, and are separated by a weak 
first order phase transition. In addition, there is an intermediate phase with 
eight peaks (whose analog may be realized in the quantum dimer model 
\cite{Ral08}), separated from the other phases by second order phase transitions
\cite{Bru75}. If one would fine-tune to these transitions, the Goldstone boson 
would become exactly massless. Even then it could not be interpreted as a dual 
photon, because the $SO(2)$ symmetry would still remain explicitly broken.
\begin{figure}[tbp]
\includegraphics[width=0.29\textwidth]{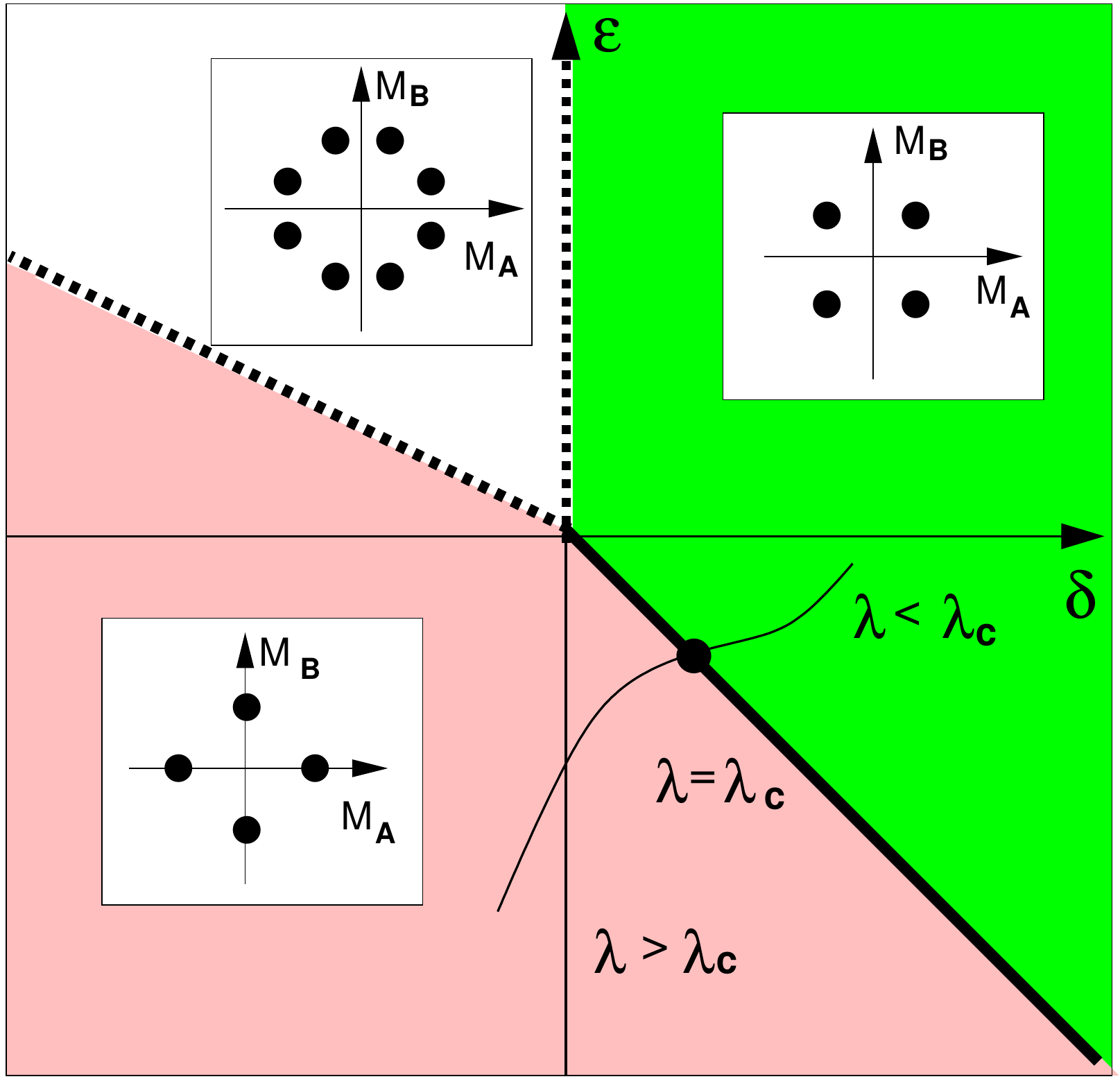}
\caption{[Color online] \textit{Phase diagram as a function of $\delta$ and 
$\varepsilon$.  The insets indicate the location of the peaks in the 
distribution $p(M_A,M_B)$. The fat and dashed lines are first and second order 
phase transitions, respectively. The curved line indicates a possible path 
taken in the QLM when varying $\lambda$.}}
\end{figure}

The effective theory predicts a finite-volume rotor spectrum 
($E_m = m^2 c^2/(2 \rho L_1 L_2)$ when $\delta = \varepsilon$ = 0) with 
$m = 0, \pm 2, \pm 4, \dots$ States with odd values of $m$ are excluded because 
they are not invariant against a sign-change of $\vec e(x)$. The quantum numbers
of the states with $m = 0, \pm 2, \pm 4$ correspond to $C = +, p = (0,0)$, 
$C = \pm, p = (\pi,\pi)$, $C = \pm, p = (0,0)$, respectively. The effective 
theory is in quantitative agreement with the exact diagonalization study
(cf.\ Fig.2b). A global fit of the energy spectrum yields 
$\lambda_c = - 0.359(5)$, $\rho = 0.45(3) J$, $c = 1.5(1) Ja$, 
$\delta_c = - \varepsilon_c = 0.01(1) J/a^2$. A more precise determination of 
the low-energy constants and of $\lambda_c$, based on high accuracy Monte Carlo 
simulations using the cluster algorithm, will be presented elsewhere.

Away from the critical point, the $\delta$-term gives rise to two distinct 
coexisting phases that are related by C and T for $\lambda < \lambda_c$, 
and by T for $\lambda > \lambda_c$. As illustrated schematically in Fig.5a 
for $\lambda \rightarrow - \infty$, the interface that separates the two phases 
represents a string of electric flux $\frac{1}{2}$. Its interface tension,
$\sigma_{1/2} = \sqrt{2|\delta|\rho}$ (for $\varepsilon = 0$), which plays the 
role of a string tension, would vanish at the phase transition if there was no 
$\varepsilon$-term. While the $\varepsilon$-term would simply be 
irrelevant at a critical point, here it is dangerously irrelevant. Taking 
it into account, the string tension never vanishes, and is always of order 
$\sqrt{\rho \varepsilon}$. Indeed, the potential (shifted by a constant) 
between two static charges $\pm 2$, illustrated in Fig.5b, shows linear 
confinement at large distances, even at the phase transition, albeit with a
small string tension $\sigma_2 = 0.201(2) J/a$ (compared to 
$\sigma_2 = 1.97(1) J/a$ at $\lambda = -1$). This shows explicitly that the 
phase transition is not a deconfined quantum critical point.
\begin{figure}[t]
\includegraphics[width=0.2\textwidth,angle=0]{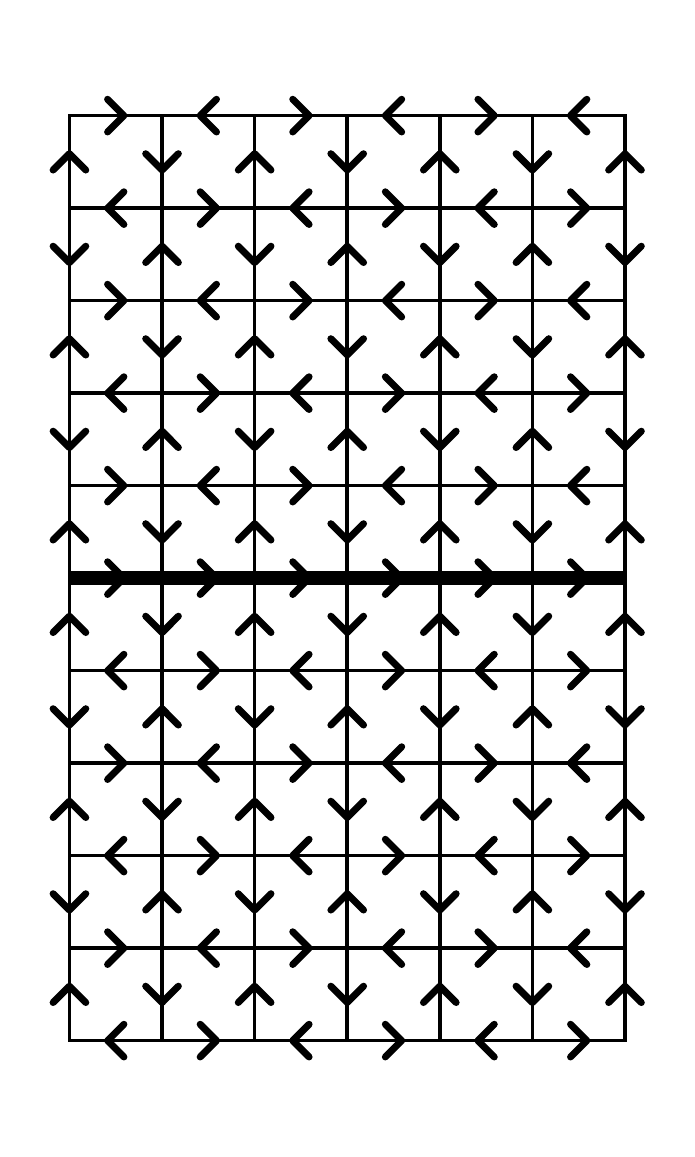}
\includegraphics[width=0.23\textwidth]{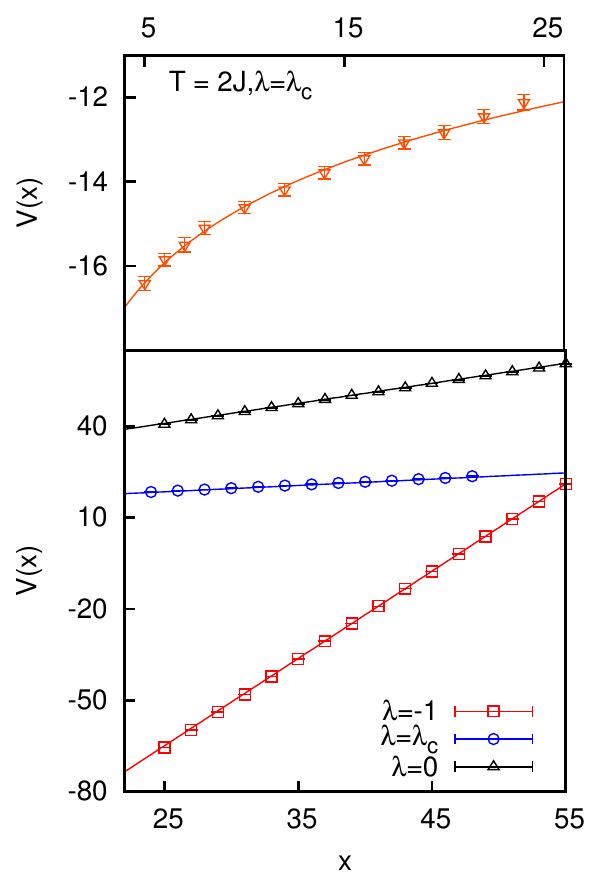} \\
\caption{[Color online] \textit{a) Interface at $\lambda \rightarrow - \infty.$
The fat line is a flux $\frac{1}{2}$ string. b) Potential between two static 
charges $\pm 2$ separated by the distance $(x,x)$ along a lattice 
diagonal, for $\lambda = -1, \lambda_c$, and 0, at $T = 0$, and at 
$\lambda = \lambda_c$ for $T = 2J$.}}
\end{figure}
The energy density $-J \langle U_\Box + U_\Box^\dagger\rangle$ in the presence of 
two charges $\pm 2$ is illustrated in Fig.6a-d. The flux string connecting the 
charges separates into four strands of flux $\frac{1}{2}$ that repel each other.
In accordance with the effective theory, the interior of the strands consists of
the phase that is stable on the other side of the transition. Near $\lambda_c$ 
the flux string undergoes topology change by wrapping one strand over the 
periodic boundary and materializing an additional strand at the edge of the 
system, whose interior then expands to become the new bulk phase (cf.\ 
Fig.\ 6b). Viewed as interfaces separating bulk phases, the strands display the 
universal phenomenon of complete wetting. 
\begin{figure}[t]
\includegraphics[width=0.5\textwidth]{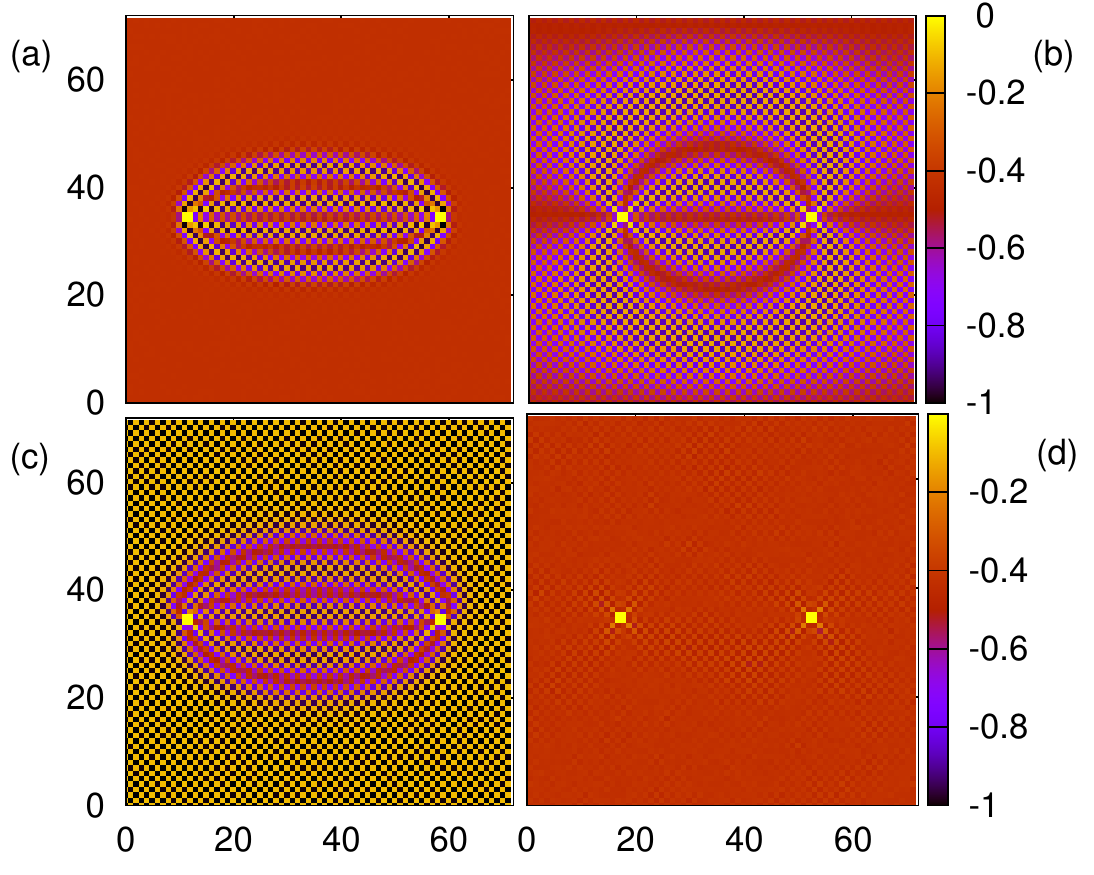}
\caption{[Color online] \textit{Energy density 
$-J \langle U_\Box + U_\Box^\dagger\rangle$ in the presence of two charges 
$\pm 2$ for $\lambda = - 1$ (a), $\lambda = \lambda_c$ (b), and $\lambda = 0$ 
(c) at $T = 0$, as well as for $\lambda = 0$ at $T > T_c$ (d).}}
\end{figure}

Finally, we have studied the system at finite temperature. The electric flux 
susceptibility, $\langle E_i^2 \rangle$ is non-zero at $T>T_c$, indicating a 
massless mode that transforms non-trivially under the $U(1)$ center symmetry, 
giving rise to a logarithmic charge-anti-charge potential (c.f.\ Fig.5b). Hence,
the ``deconfined'' phase no longer has linear, but still has logarithmic 
confinement. As illustrated in Fig.6d, the flux then spreads out and no longer 
forms a string. Interestingly, the shift symmetry T remains spontaneously broken
at high temperature (c.f.\ Fig.3d). Actually, in the deconfined phase yet 
another $SO(2)$ symmetry emerges, which originates from the Gauss law.

In conclusion, we have observed an emergent $SO(2)$ symmetry with an associated
pseudo-Goldstone boson in the $(2+1)$-d $U(1)$ QLM. Interfaces
separating phases with spontaneously broken C or T symmetry manifest
themselves as strings carrying fractional electric flux $\frac{1}{2}$. Although 
the model displays certain features of deconfined quantum critical points,
a dangerously irrelevant operator leads to small explicit $SO(2)$ breaking.
This prevents the interpretation of the emergent Goldstone boson as a massless 
photon, and implies a non-zero string tension also at the phase transition.
It remains to be seen whether phenomena, similar to the ones observed in the 
QLM, may masquerade as deconfined quantum criticality in other models as well. 
Once the $(2+1)$-d $U(1)$ QLM is realized in ultracold matter experiments, its 
rich dynamics will become accessible to quantum simulation.

We dedicate this work to the memory of Bernard B.\ Beard. He is deeply missed,
not only as a collaborator, who would have elegantly performed all simulations 
in this paper in continuous time \cite{Bea96}. We like to thank W.\ Bietenholz, 
M.\ Dalmonte, C.\ P.\ Hofmann, E.\ Katz, A.\ L\"auchli, M.\ L\"uscher,
F.\ Niedermayer, G.\ Palma, E.\ Rico, A.\ Sen, and P.\ Zoller for illuminating 
discussions. UJW acknowledges B.\ B.\ Beard, R.\ Brower, S.\ Chandrasekharan, 
V.\ Chudnovsky, U.\ Gerber, M.\ Pepe, and A.\ Tsapalis for their collaboration 
on previous attempts to simulate quantum links. This research has been supported
by the 
Schweizerischer Nationalfonds.

\end{document}